\documentclass[aps,prb,twocolumn,superscriptaddress,amsmath,amssymb]{revtex4}
\usepackage{mathrsfs}
\usepackage{graphicx}
\usepackage{dcolumn}
\usepackage{bm}
\usepackage{amsmath}
\usepackage{amsfonts}

\begin{document}

\title{Singlet Mott State Simulating the Bosonic Laughlin Wave Function}
\author{Biao Lian}
\affiliation{Department of Physics, McCullough Building, Stanford University, Stanford, California 94305-4045, USA}
\author{Shoucheng Zhang}
\affiliation{Department of Physics, McCullough Building, Stanford University, Stanford, California 94305-4045, USA}
\date{\today}

\begin{abstract}
We study properties of a class of spin singlet Mott states for arbitrary spin $S$ bosons on a lattice, with particle number per cite $n=S/l+1$, where $l$ is a positive integer. We show that such a singlet Mott state can be mapped to a bosonic Laughlin wave function on the sphere with a finite number of particles at filling $\nu=1/2l$. Spin, particle and hole excitations in the Mott state are discussed, among which the hole excitation can be mapped to the quasi-hole of the bosonic Laughlin wave function. We show that this singlet Mott state can be realized in a cold atom system on optical lattice, and can be identified using Bragg spectroscopy and Stern-Gerlach techniques. This class of singlet Mott states may be generalized to map to bosonic Laughlin states with filling $\nu=q/2l$.
\end{abstract}
\maketitle
While the spin $1/2$ electron systems have been extensively studied in the past decades, much of the physics of large spin particle systems is still unexplored today, because of the lack of such quantum systems in nature. With the development of ultracold atoms on optical lattice, the realization of such systems becomes accessible. Recently \cite{McClelland2006,Aikawa2012,Lu2011}, the Bose-Einstein condensates (BEC) of Er atoms with spin $6$ and Dy atoms with spin $8$ are realized at temperatures as low as $10\sim10^2$ nK. The spins of both atoms originate from the spin and orbital angular momenta of the outermost shell electrons, with vanishing nuclear spins. Theoretically, the condensates of large spin bosons have been shown to have various novel phases which support non-Abelian superfluid vortices \cite{Ho1998,Ohmi1998,Yip2007,Barnett2006,Barnett2009,Lamacraft2010,Kawaguchi2011,Lian2012,Mermin1979}. In principle, when a background optical lattice is introduced, many more interesting quantum states such as the AKLT state \cite{Affleck1987} may arise in the Mott regime. So far, most studies on large spin Mott states have been focused on spin 1 and 2 bosons \cite{Demler2002,Hou2003,Tsuchiya2004,Krutitsky2004,Krutitsky2005,Pai2008,Snoek2009} or spin $3/2$ fermions \cite{Wu2003,Wu2005}. The Mott physics of the larger spin particles is basically unknown and remains to be studied. In this letter, we study the spin singlet Mott states of bosons of arbitrary spin, and give a simple physical picture. In the large spin limit, these states simulate the bosonic Laughlin wave function \cite{Laughlin1983,Haldane1983}. Whereas fermionic Laughlin wave function is realized by electrons in the fractional quantum Hall state, the bosonic Laughlin wave function has not been realized in nature before our work.

The main idea is the exact mapping between the particle spin and the particle coordinates on a sphere\cite{Arovas1988}. Based on this idea, we show that for spin $S$ bosons in an optical lattice, there are a class of singlet Mott states which can be exactly mapped to the bosonic Laughlin wave function. Particularly, large spin bosons Er or Dy may be used as a suitable system to `realize' a bosonic Laughlin state with a finite number of particles, which is comparable to the particle number used in current numerical calculations\cite{Murthy2003}.
This letter is organized as follows. First, we construct the spin Bose-Hubbard model for spin $S$ bosons in an optical lattice, and demonstrate that under certain constraints of the Hubbard interactions it has a spin singlet Mott insulator ground state. We show such a Mott state can be exactly mapped to the $\nu=1/2l$ filling bosonic Laughlin wave function on the sphere, where $l$ is a positive integer. The experimental realization of the singlet Mott state is then discussed. We study the properties of excitations of this state and give the criteria for identifying the singlet Mott state in the cold atom experiments. Finally, we briefly discuss the generalization to singlet Mott states corresponding to the higher $\nu=q/2l$ filling bosonic fractional quantum Hall states.


Without loss of generality, the spin $S$ bosons on an optical lattice can be well described by the spin Bose-Hubbard model Hamiltonian $H=H_I+H_t$, with
\begin{equation}\label{BH}
\begin{split}
&H_I=-\mu\sum_{i} \hat{n}_i+\frac{1}{2}\sum_i\Big[U\hat{n}_i(\hat{n}_i-1) +\sum_{J>0}^{2S}U_J\hat{\mathcal{P}}^S_J(i)\Big],\\
&H_t=-\sum_{\langle ij\rangle,m}\left(t\psi_{i,m}^\dag\psi_{j,m}+h.c.\right) \ ,
\end{split}
\end{equation}
where $t$ is the nearest neighbour hopping amplitude, $\mu$ is the chemical potential, while $U$ and $U_J$ are the on-site Hubbard interaction parameters. $i,j$ are the lattice site indices, and $m$ is the spin $z$ component index. $\hat{n}_i=\sum_m\psi_{i,m}^\dag\psi_{i,m}$ is the particle number on site $i$. The projection operator $\hat{\mathcal{P}}^S_J(i)$ are non-negative operators defined by $\hat{\mathcal{P}}^S_J(i)=\sum_m\mathcal{A}^\dag_{Jm}(i)\mathcal{A}_{Jm}(i)$ in terms of $\mathcal{A}_{Jm}(i)=\sum_{m'}\langle Jm|Sm',Sm-m'\rangle\psi_{i,m'}\psi_{i,m-m'}$, where $\langle Jm|Sm',Sm-m'\rangle$ is the Clebsch-Gordan coefficient. It projects the two-particle on-site interaction into the total spin $J$ channel, and satisfies the identity $\sum_J\hat{\mathcal{P}}^S_J(i)=\hat{n}_i(\hat{n}_i-1)$. For bosons, $J$ is restricted to even integers by symmetry requirements. To avoid the collapse instability of bosons, we require $U>0$ and $U+U_J>0$ for any $J$.

In the Mott regime where $|t|\ll U$ \& $U_J$, we can treat the hopping term $H_t$ as the perturbation, and to the lowest order set $H=H_I$. Under this approximation, all the lattice sites are decoupled, and the ground state $|\Psi\rangle$ is simply the direct product of the lowest on-site states $|\varphi\rangle_i$:
\begin{equation}
|\Psi\rangle=\prod_i|\varphi\rangle_i\ .
\end{equation}
Since $H_I$ has the U($1$) charge symmetry and SU($2$) spin rotation symmetry, the on-site state $|\varphi\rangle_i$ is the eigenstate of the on-site particle number $\hat{n}_i$ and the on-site total spin $\hat{\mathbf{S}}^2_i=(\psi_{i,a}^\dag\mathbf{S}_{ab}\psi_{i,b})^2$, where $\mathbf{S}_{ab}$ is the representation matrix of spin $S$. Namely, if there are $n$ particles on site $i$, their spins will combine into a definite total spin.

To clearly see the composition of different particle spins on the same site, we use the Schwinger boson representation for the spin of each particle. The spin of the $\alpha$-th particle ($\alpha$ runs from $1$ to $n$) on lattice site $i$ is denoted by $\mathbf{s}_{i\alpha}$, and is defined as:
\begin{equation}\label{Schwinger}
\begin{split}
&s^+_{i\alpha}=a^\dag_{i\alpha} b_{i\alpha},\qquad s^z_{i\alpha}=\frac{1}{2}(a^\dag_{i\alpha}a_{i\alpha}-b^\dag_{i\alpha}b_{i\alpha}),\\
&s^-_{i\alpha}=a_{i\alpha}b^\dag_{i\alpha},\qquad s^{\text{tot}}_{i\alpha}=\frac{1}{2}(a^\dag_{i\alpha}a_{i\alpha}+b^\dag_{i\alpha}b_{i\alpha}),
\end{split}
\end{equation}
where $s^\pm_{i\alpha}=s^x_{i\alpha}\pm is^y_{i\alpha}$ are the raising and lowering operators in the spin algebra. $a_{i\alpha}$ and $b_{i\alpha}$ are the commonly defined bosonic annihilation operators. For the particle to have spin $S$, the physical state $|\varphi\rangle_i$ is required to satisfy $s^{\text{tot}}_{i\alpha}|\varphi\rangle_i=S|\varphi\rangle_i$.

We now investigate the simple case when the spin-dependent Hubbard interaction
\begin{equation}\label{condition}
\begin{split}
&U_J\ge0\qquad \text{for}\qquad J>2S-2l\ ,\\
&U_J=0\qquad \text{for} \qquad J\le2S-2l\ ,
\end{split}
\end{equation}
where integer $l$ is a factor of $S$. In this case, any two particles on the same site prefer to have their composite spin $J$ no higher than $2S-2l$, so that the spin-dependent Hubbard interaction energy can be minimized to zero. If the chemical potential $\mu$ lies in the interval $U(S/l)<\mu<U(S/l+1)$, the following spin singlet state becomes the on-site ground state:
\begin{equation}\label{Mott}
|\varphi\rangle_i=\prod_{1=\alpha<\beta}^{n}\left(a^\dag_{i\alpha}b^\dag_{i\beta}-b^\dag_{i\alpha}a^\dag_{i\beta}\right)^{2l}|0\rangle_i\ ,
\end{equation}
where $n=S/l+1$ is the particle number of site $i$, while $|0\rangle_i$ is the vacuum for the Schwinger bosons $a_{i\alpha}$ and $b_{i\alpha}$. The even power $2l$ outside each bracket ensures the bosonic nature of the state. This expression is a homogeneous polynomial of $a_{i\alpha}$ and $b_{i\alpha}$ of degree $2S$, so $|\varphi\rangle_i$ indeed describes spin $S$ of each particle. Since any two particle spins $\mathbf{s}_{i\alpha}$ and $\mathbf{s}_{i\beta}$ in state $|\varphi\rangle_i$ are coupled by $2l$ singlet Schwinger pairs, their composite spin $\mathbf{J}_i(\alpha,\beta)=\mathbf{s}_{i\alpha}+\mathbf{s}_{i\beta}$ cannot exceed $2S-2l$, and they will have zero projection on $\hat{\mathcal{P}}^S_J(i)$ for $J>2S-2l$. Therefore, the single-site energy of this state is minimized to $\mathcal{E}_{g}(i)=-\mu n+Un(n-1)/2$. On the other hand, since all the Schwinger bosons are paired into singlets, the entire state is naturally a spin singlet state, namely $\hat{\mathbf{S}}_i^2|\varphi\rangle_i=(\sum_\alpha\mathbf{s}_{i\alpha})^2|\varphi\rangle_i=0$. This Mott state thus does not break any symmetry.

We note that in analogy to the argument of Arovas, Auerbach and Haldane \cite{Arovas1988}, the on-site state $|\varphi\rangle_i$ can be exactly mapped to a bosonic Laughlin wave function on the sphere. To see this, we can rewrite the state $|\varphi\rangle_i$ under the spin coherent state basis. This is most easily done by taking $a_{i\alpha}^\dag\rightarrow u_{i\alpha}$, $b_{i\alpha}^\dag\rightarrow v_{i\alpha}$, $a_{i\alpha}\rightarrow \partial/\partial u_{i\alpha}$, $b_{i\alpha}\rightarrow \partial/\partial v_{i\alpha}$, where $u_{i\alpha}=\cos(\theta_{i\alpha}/2)e^{i\phi_{i\alpha}/2}$, $v_{i\alpha}=\sin(\theta_{i\alpha}/2)e^{-i\phi_{i\alpha}/2}$ labels the spin coherent state of the $\alpha$-th particle pointing along the unit sphere coordinates $(\theta_{i\alpha},\phi_{i\alpha})$ \cite{Arovas1988}. The resulting wave function is
\begin{equation}
\varphi_i=\prod_\alpha(u_{i\alpha}^\dag,v_{i\alpha}^\dag)|\varphi\rangle_i =\prod_{1=\alpha<\beta}^{n}\left(u_{i\alpha}v_{i\beta}-v_{i\alpha}u_{i\beta}\right)^{2l}\ .
\end{equation}
This total symmetric function is derived by Haldane as the $\nu=1/2l$ filling $n$-body Laughlin wave function on the sphere\cite{Haldane1983}, which in the thermodynamic limit describes a bosonic fractional quantum hall state with a fractional Hall conductance $\sigma_{xy}=\nu e^2/h$. The effective ``magnetic flux" through the whole sphere is $\Phi=2S$, and the filling condition $n-1=\nu\Phi$ is automatically satisfied.

The bosonic quantum hall states have been studied theoretically for strongly correlated bosonic systems \cite{Cooper2001,Senthil2013}, but are still challenging for experimental studies. It will be interesting if these states can be experimentally realized in large spin atomic systems, though in the sense of mapping equivalence. For Dy atoms which have spin $S=8$ and for the filling fraction $\nu=1/2$, the number of particles per cite is $n=9$. Though far from the thermodynamic limit, this number of particles is comparable to that adopted in the recent exact diagnolization calculations \cite{Murthy2003}.

In cold atom systems, the Hubbard interaction parameters can be tuned through resonance methods. To the lowest order, they are given by $U=4\pi\hbar^2a_0/M$, and $U_J=4\pi\hbar^2(a_J-a_0)/M$, where $a_J$ is the two-particle $s$-wave scattering length in the total spin $J$ channel, and $M$ is the atomic mass. In the strongly interacting regime, the shape of the optical lattice potential can also alter the parameters $U$ and $U_J$ significantly. By using the Feshbach resonance and shape resonance methods commonly used in cold atom experiments, it is promising to reach the regime where $U_{2S}$ is positive and near resonance, and is therefore far greater than all the other $U_J$. According to Eq. (\ref{condition}), this is the condition for realizing the $l=1$ spin singlet Mott state, which corresponds to the mapping onto the $\nu=1/2$ filling bosonic Laughlin wave function. Provided the anisotropy introduced by the resonance methods is small enough, the spin singlet Mott state will stay robust.

We now turn to the excitation properties of the state. Since the Mott state does not break any symmetry, there is no Goldstone mode and all the excitations in the system are gapped. To discuss the energy-momentum dispersion of the excitations, we take into account the perturbative hopping term $H_t$ introduced in Eq. (\ref{BH}).

The first type of excitation is the particle or hole excitation. A static particle or hole excitation localized on the lattice site $i$ is represented by a state $|i,\pm,\Psi\rangle=|\varphi^\pm\rangle_i\otimes\prod_{j\neq i}|\varphi\rangle_j$, where $+$ and $-$ correspond to particle and hole, respectively. The on-site states $|\varphi^\pm\rangle_i$ are obtained by acting $(\psi^\dag_i\cdot\xi)$ and $(\xi^\dag\cdot\psi_i)$ on state $|\varphi\rangle_i$, where $\xi$ stands for a spinor of $2S+1$ components. Both excitations carry spin $S$. Particularly, under the spin-coherent state basis, the wave function of the hole state $|\varphi^-\rangle_i$ can be written in a general form
\begin{equation}
\begin{split}
&\varphi^-_i=\prod_{\sigma=1}^{2l}\left[\prod_{\gamma=1}^{n-1}\left(v_{\xi,\sigma} u_{i\gamma}-u_{\xi,\sigma} v_{i\gamma}\right)\right] \\ &\qquad\qquad\qquad\times\prod_{1=\alpha<\beta}^{n-1}\left(u_{i\alpha}v_{i\beta}-v_{i\alpha}u_{i\beta}\right)^{2l}\ ,
\end{split}
\end{equation}
where $(u_{\xi,\sigma},v_{\xi,\sigma})$ are fixed $2$-component spinors. We note that this wave function resembles the quasi-hole excitation wave functions on the $\nu=1/2l$ filling bosonic Laughlin state \cite{Laughlin1983}. The $2l$ quasi-holes are separately distributed on the sphere, whose coordinates are in accordance with the directions of the $2l$ spinors $(u_{\xi,\sigma},v_{\xi,\sigma})$. The energy gap of the hole excitation can be determined to be $\Delta_-=\mu-U(n-1)\sim U$, while that of the particle excitation $\Delta_+\sim U+\sum_J U_{J}$.

\begin{figure}
\includegraphics[width=3.0in]{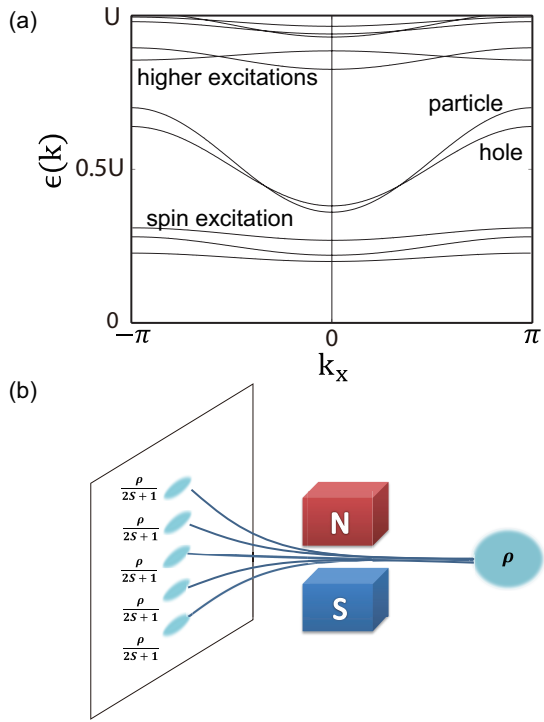}
\caption{Illustration of the criteria for experimentally identifying the spin singlet Mott state. (a) An illustrative energy spectrum of the excitations. The vertical and horizontal axes are the energy in unit of $U$ and the $x$ direction momentum $k_x$ respectively. In the Mott regime, the band widths of the spin excitations are much smaller than that of the particle and hole excitations. (b) Stern-Gerlach measurement of the singlet Mott state. For illustration purposes, the spin is set to $S=2$. The populations on all $z$-direction spin components are equal to each other due to the spin rotational invariance. \label{Stern-Gerlach}}
\end{figure}

The hopping term $H_t$ can induce the motion of the particle and hole excitations, and can also lead to a fusion of them into other excitations. In the spirit of the single-mode approximation (SMA), we can use $|\mathbf{k},\pm,\Psi\rangle= (1/\sqrt{N})\sum_ie^{i\mathbf{k}\cdot\mathbf{R}_i}|i,\pm,\Psi\rangle$ as a variational wave function, and use $\epsilon_\pm({\mathbf{k}})=\langle \mathbf{k},\pm,\Psi|H-E_g|\mathbf{k},\pm,\Psi\rangle$ to determine the dispersion of the particle and the hole, where $E_g=\sum_i\mathcal{E}_g(i)$ is the ground state energy. For square lattice for instance, they are given by
\begin{equation}
\epsilon_\pm({\mathbf{k}})=\Delta_\pm-\frac{(2n+1\pm1)t}{2S+1}(\cos k_x+\cos k_y)\ .
\end{equation}
We note the $k$-dependence of the hole derived here agrees with that derived for the single hole motion in the AKLT state \cite{Zhang1989}, where $n=1$ and $S=0$ or $1$.

Another type of excitation is the spin excitaion, which carries a spin but keeps the particle number in the state invariant. The local state for a pure spin excitation can be represented as $|i,J_S,\Psi\rangle=|\varphi^{J_S}\rangle_i\otimes\prod_{j\neq i}|\varphi\rangle_j$, where $|\varphi^{J_S}\rangle_i$ is an $n$-particle state on the $i$-th lattice site satisfying $\hat{\mathbf{S}}_i^2|\varphi^{J_S}\rangle_i=J_S(J_S+1)|\varphi^{J_S}\rangle_i$. The simplest spin excitation state can be written by acting the particle-hole pair operator $\sum_{m_1}(-1)^{m_1}\langle J_Sm|Sm_1,Sm-m_1 \rangle\psi^\dag_{i,-m_1}\psi_{i,m-m_1}$ on the state $|\varphi\rangle_i$. The gap of a spin excitation can be estimated to be $\Delta_{J_S}\sim \sum_J U_J$. Thus for the case the spin-independent Hubbard parameter $U$ is large, these spin excitations are the low energy excitations.

We can again use the SMA to estimate the dispersion of spin excitations, assuming the variational wave function $|\mathbf{k},J_S,\Psi\rangle= (1/\sqrt{N})\sum_ie^{i\mathbf{k}\cdot\mathbf{R}_i}|i,J_S,\Psi\rangle$ for a moving spin. Since $H_t$ changes the on-site particle numbers and is not closed for spin excitations, the leading contribution to the energy-momentum dispersion comes from the $2$nd order perturbation, $\epsilon_{J_S}(\mathbf{k})=\langle \mathbf{k},J_S,\Psi|(H_I-E_g)-H_t(H_I-E_g)^{-1}H_t|\mathbf{k},J_S,\Psi\rangle$. This gives us a spin dispersion
\begin{equation}
\epsilon_{J_S}(\mathbf{k})\sim\Delta_{J_S}-\frac{n^2t^2}{(2J_S+1)^2U}(\cos k_x+\cos k_y)\ .
\end{equation}
A spin excitation is thus much less mobile than a particle or hole excitation.

We have shown in Fig. \ref{Stern-Gerlach}(a) an illustrative energy spectrum of excitations, in which one can easily identify the particle and hole excitations by comparing the band widths. Due to the effect of hopping, a spin excitation can split into a particle and a hole, and in accordance a particle and a hole can also combine into a pure spin. Via higher order processes, they can in principle form many more complex excitations with higher energies, about which we will not discuss here. In general, the fusion and combination processes of the excitations will give them a finite life time.

We can give the criteria for experimentally identifying such a spin singlet Mott state in an optical lattice cold atom system. In cold atom experiments, the energy-momentum dispersions of excitations in the system can be measured with the Bragg spectroscopy \cite{Stamper-Kurn1999}. One can also measure the populations of the atoms on each $z$-direction spin component by the Stern-Gerlach technique combined with the light absorption method. These two measurements are sufficient to test the criteria.

First, for the spin singlet Mott state, the energy spectrum of excitations measured should be all gapped, as is shown in the illustrative example in Fig. \ref{Stern-Gerlach}(a). This is different from that of the symmetry breaking states with a spin order or a U($1$) phase order, which will have gapless Goldstone mode excitations (magnons and phonons). Further, we can also distinguish between the spin singlet Mott state here and a gapped AKLT state, by looking at the energy gap of the excitations. In the spin singlet Mott state $|\Psi\rangle$, all the excitation gaps are of order $\Delta_{(\pm,J_S)}\sim U$, and depends little on the hopping amplitude $t$. In the AKLT state, the excitation gap is proportional to the superexchange spin-spin interaction \cite{Affleck1987}, $\Delta_{AKLT}\sim t^2/U$. Through a variation of the parameter $U$ or $t$, one can easily distinguish between these two states by examining the change of the excitation gap.

Second, the SU($2$) spin rotational invariance of the spin singlet Mott state can be verified by a Stern-Gerlach measurement. This technique has been widely used in cold atom experiments. It is easy to show that due to the spin rotational invariance, the population per volume of particles on each $z$-direction spin component is given by $\rho_m=\langle\Psi|\sum_i \psi^\dag_{i,m}\psi_{i,m}|\Psi\rangle/V=\rho/(2S+1)$ independent of $m$, where $\rho=\langle\Psi|\sum_i \hat{n}_i|\Psi\rangle/V$ is the total density. The result does not change if the Stern-Gerlach measurement is carried out along another direction. This is illustrated in Fig. \ref{Stern-Gerlach}(b). For any spin ordered Mott states or spinor Bose-Einstein condensate states, the SU($2$) symmetry is broken, and the populations on different components will depend on choice of the direction of measurement. These two steps form the criteria for identifying our proposed spin singlet Mott state in the cold atom experiment.

We now briefly discuss the generalized spin singlet Mott states simulating a $\nu=q/2l$ filling bosonic fractional quantum Hall state in the large $S$ limit, where $q$ is a positive integer. Such a state $|\Psi(q)\rangle$ may be written down as $|\Psi(q)\rangle=\prod_i|\varphi(q)\rangle_i$, where the on-site state $|\varphi(q)\rangle_i$ consists of $q$ multiples of the $n$-particle spin singlet defined in Eq. (\ref{Mott}), where $n=S/l+1$. So this state is naturally a spin singlet state. The particle number per site of this state is $q n$. Under the spin-coherent state basis, the wave function of $|\varphi(q)\rangle_i$ takes the form
\begin{equation}
\varphi(q)_i=\sum_{\mathcal{P}}\prod_{r=0}^{q-1} \prod_{nr+1=\alpha<\beta}^{n(r+1)}\left(u_{i\mathcal{P}\alpha}v_{i\mathcal{P}\beta} -v_{i\mathcal{P}\alpha}u_{i\mathcal{P}\beta}\right)^{2l}\ ,
\end{equation}
where $\mathcal{P}$ runs over all the permutations of the particle indices. This wave function is a generalization of the $\nu=1/2l$ filling Laughlin wave function on the sphere. With a suitable choice of the Hubbard parameters $U$ and $U_J$ and the chemical potential $\mu$, such a generalized spin singlet Mott state may be preferred as a ground state. The excitation gaps of the state $|\Psi(q)\rangle$ are still of order $U$, but the band energy widths of the particle (hole) excitation and the spin excitation will be broadened by a factor $q$ and $q^2$ respectively.

In summary, we studied the conditions for realizing the spin singlet Mott states for arbitrary spin bosons in the lattice, which may be met through the resonance methods in a cold atom system. Such singlet Mott states can be mapped to the $\nu=1/2l$ filling bosonic Laughlin wave functions. Various excitations like the particle, hole and spin excitations arise, which all have finite gaps. The hole excitation resembles the quasi-holes in the bosonic Laughlin state. We proposed the criteria for identifying the spin singlet Mott state in the cold atom experiment, which make use of the Bragg spectroscopy and the Stern-Gerlach techniques. Lastly, we made a short discussion on the generalized singlet Mott states, which can be mapped to the $\nu=q/2l$ filling Laughlin wave functions.

\emph{Acknowledgements.} This work is supported by the NSF under grant numbers DMR-1305677.

\bibliography{HighS-ref}

\end{document}